\documentclass[twocolumn,showpacs,preprintnumbers,amsmath,amssymb]{revtex4}

\usepackage{graphicx}
\usepackage{dcolumn}
\usepackage{bm}
\usepackage{latexsym}
\usepackage{amsfonts}
\usepackage{amssymb}
\usepackage{amsmath}

\begin{document}


\title{  Lorentz Violating Inflation  }

\author{Sugumi Kanno}
\email{sugumi@hep.physics.mcgill.ca}
\affiliation{
 Department of Physics,  McGill University, Montr${\acute e}$al, 
 QC H3A 2T8, Canada
}%
\author{Jiro Soda}
\email{jiro@tap.scphys.kyoto-u.ac.jp}
\affiliation{
 Department of Physics,  Kyoto University, Kyoto 606-8501, Japan
}%

\date{\today}

\begin{abstract}
 We explore the impact of Lorentz violation on the inflationary scenario.
 More precisely, we study the inflationary scenario 
  in the scalar-vector-tensor theory
 where the vector is constrained to be unit and time like.
 It turns out that the Lorentz violating vector affects the dynamics of the
 chaotic inflationary model and divides the inflationary stage into
 two parts; the Lorentz violating stage and the standard slow roll
 stage. We show that the universe is expanding as an exact de Sitter
 spacetime in the Lorentz violating stage although the inflaton field
 is rolling down the potential. Much more interestingly, 
 we find exact Lorentz violating inflationary solutions
 in the absence of the inflaton potential. In this case, the inflation
 is completely associated with the Lorentz violation. 
 We also mention some consequences
 of Lorentz violating inflation which can be tested by observations. 
\end{abstract}

\pacs{98.80.Cq, 98.80.Hw}
\maketitle

\section{Introduction}

The Lorentz invariance has been considered as the most fundamental
symmetry of physics. However, as far as we know, any symmetry
is not realized exactly or spontaneously broken. 
 Hence, it is important to investigate a possibility
of violation of Lorentz invariance. In fact,
the observation of high energy cosmic rays reports super GZK events
though it needs further confirmation~\cite{Greisen,
Zatsepin:1966jv,Takeda:1998ps}.
This may suggests the Lorentz violation~\cite{Sato:2000vu,Coleman:1998ti}. 
Theoretically, any quantum theory of gravity requires drastic modification
of the picture of space-time at the Planck scale~\cite{Amelino-Camelia:1997gz}. 
Some of them yields the Lorentz violation~\cite{Kostelecky:1988zi}.

The impact of Lorentz violation on physics should be broad.
Even in the cosmology, there are many subjects which should be reconsidered.
For example, the dark matter and the dark energy problem, the inflation,
the baryogenesis, cosmic rays, 
and so on~\cite{Bekenstein:2004ne,Carroll:2004ai,Carroll:2005dj}.
Here, we shall concentrate on the inflationary scenario.

Typically, the Lorentz violation yields the preferred frame.
In the case of the standard model of particles, 
there are strong constraints on the 
existence of the preferred frame~\cite{Mattingly:2005re}. 
In contrast, there is no reason to refuse the preferred frame in cosmology.
Rather, there is a natural preferred frame
which is defined by cosmic microwave background radiation (CMB).
Therefore, there is room to consider the gravitational theory
which allows the preferred frame. 
The purpose of this paper is to clarify what occurs in inflationary stage 
when we allow the preferred frame from the beginning. 
 It turns out that there is a chance
to detect the evidence of the Lorentz violation through the observation of
the cosmic microwave background radiation and the primordial 
gravitational waves.

When we talk about the Lorentz violation, 
we have to specify the model somehow.
Recently, various types of theory of gravity with the Lorentz violation 
are proposed. One is the ghost condensation model which has a non-conventional
kinetic term. In the stable vacuum, the kinetic term has the expectation value. 
Hence, the Lorentz invariance is violated spontaneously~\cite{Arkani-Hamed:2003uy,Arkani-Hamed:2004ar,Cheng:2006us}.
This violation mechanism is an interesting possibility.
The inflationary scenario in the ghost condensation was also investigated
in this context~\cite{Arkani-Hamed:2003uz}.  Another interesting one is 
brane model~\cite{Csaki:2000dm,Cline:2003xy,Libanov:2005nv}. 
 From the string theoretical point of view, the braneworld picture seems to
 be natural. Hence, it is important to examine the Lorentz violation
 in the braneworld. 
 Of course, there are other interesting models.

In this paper, we will consider the spontaneous breaking of
Lorentz symmetry due to a vector field~\cite{Gripaios:2004ms}.
 When this vector field couples to the gravity, we obtain
 the Lorentz violating theory of gravity, 
 the so-called Einstein-Ather theory~\cite{Jacobson:2000xp}. 
Interestingly, this theory has a wide parameter region
where all of current experiments and observations can be 
explained~\cite{Eling:2003rd,Graesser:2005bg,Foster:2005dk,Foster:2006az}.
When we consider the cosmology, parameters could depend on time.
The time evolution of parameters can be regarded 
as a consequence of the dynamics of a scalar field. 
Thus, a natural generalization of the Einstein-Ather theory
is the scalar-vector-tensor 
theory of gravity with a timelike unit vector field. 

Recently, Lim has studied the inflationary scenario in the 
context of the Einstein-Ather theory~\cite{Lim:2004js}.
In this paper, we will reconsider the inflationary scenario
based on the Lorentz violating scalar-vector-tensor theory of gravity.
In particular, the coupling between the inflaton and the Lorentz violating
vector is incorporated in our model. 
Our primary concern is how the Lorentz violation can affect the inflationary
scenario when we include this coupling. 
First, we show how the chaotic inflationary scenario is affected 
by the Lorentz violation. In the conventional theory of gravity,
there is a power law inflation model which is an exact solution
 with the exponential potential. Hence, it is legitimate
 to  seek for exact solutions also
 in the Lorentz violating scalar-vector-tensor theory of gravity.
 Indeed, we find the three kind of exact solutions
 in the absence of the inflaton potential. 
We also discuss the observability of Lorentz violation.

The organization of this paper is the following:
in sec II, we introduce the scalar-vector-tensor theory
 where the Lorentz symmetry is spontaneously broken due to the unit-norm
 vector field. 
 In sec. III, we study Lorentz violating chaotic inflation.
 In sec. IV, we examine the model without the inflaton potential and
 find the exact inflationary solutions. 
In sec. V, the cosmological tensor perturbations are discussed.
Final section is devoted to the conclusion.
In the Appendix, we demonstrate the alignment of two preferred frames, 
namely, the cosmological and the vector frame.

\section{Lorentz Violating Scalar-Vector-Tensor Theory}

In this section, we present our model with which we discuss
the inflationary scenario. 

We assume there exists the Lorentz symmetry but it is spontaneously
broken by getting the expectation values of a vector field $u^\mu$ as
\begin{eqnarray}
     <0| u^\mu u_\mu |0> = -1 \ .
\end{eqnarray}
The mechanism which gives this expectation value
is discussed in Ref. \cite{Kostelecky:1988zi}. 
Here, we have chosen time-like expectation value for the reason
explained in Ref. \cite{Elliott:2005va}. 
Nambu-Goldstone modes can be represented by
\begin{eqnarray}
  u^\mu = \frac{1}{\sqrt{1-\bm{\psi}^2}} 
          \left( 1, \bm{\psi} \right) \ ,
\end{eqnarray}
where $\bm{\psi}$ is a spatial vector field.  
Now, the action for the Nambu-Goldstone boson in the curved spacetime
becomes 
\begin{eqnarray}
  S &=& \int d^4 x \sqrt{-g} \left[ 
  - \beta_1 \nabla^\mu u^\nu \nabla_\mu u_\nu 
   -\beta_2 \nabla^\mu u^\nu \nabla_\nu u_\mu  \right. \\
 && \left.  -\beta_3 \left( \nabla_\mu u^\mu \right)^2
  -\beta_4 u^\mu u^\nu \nabla_\mu u^\alpha \nabla_\nu u_\alpha 
  + \lambda \left( u^\mu u_\mu +1 \right)
                             \right]  \ , \nonumber
\end{eqnarray}
where $\beta_i$ are arbitrary parameters. 
Here we have take into account the expectation value
 by just adopting it as a constraint
\begin{eqnarray}
  u^\mu u_\mu = -1 .
\end{eqnarray}
Thanks to the constraint, 
this is the most general low energy action which has derivatives up to
the second order.
Note that we take $u^\mu$ as the dimensionless vector. 
Hence, each $\beta_i$ has dimension of mass squared. 
In other words, $\sqrt{\beta_i}$
gives the mass scale of symmetry breakdown.

It is straightforward to couple this Nambu-Goldstone modes to gravity
by just adding the Einstein-Hilbert term.
The resultant theory is called Einstein-Ather or vector-tensor theory.
Here, we adopt the latter name. 
Remarkably, this vector-tensor theory is in agreement with current 
experiments as far as certain relations of parameters 
hold~\cite{Eling:2003rd,Graesser:2005bg,Foster:2005dk,Foster:2006az}.

We will consider the inflationary scenario in this Lorentz violating
gravity. Now, it is possible that the inflaton couples to
the vector in the following way 
\begin{eqnarray}
  S &=& \int d^4 x \sqrt{-g} \left[ \frac{1}{16\pi G} R
  - \beta_1 (\phi) \nabla^\mu u^\nu \nabla_\mu u_\nu \right. \nonumber\\
&& \left.  \qquad\qquad\qquad
   -\beta_2 (\phi )\nabla^\mu u^\nu \nabla_\nu u_\mu    
   -\beta_3 (\phi) \left( \nabla_\mu u^\mu \right)^2 \right. \nonumber\\
&& \left. \qquad\qquad\qquad
  -\beta_4 (\phi) u^\mu u^\nu \nabla_\mu u^\alpha \nabla_\nu u_\alpha 
                                   \right. \nonumber \\
&& \left.  \qquad\quad\quad + \lambda \left( u^\mu u_\mu +1 \right)
  - \frac{1}{2} \left(\nabla \phi \right)^2 - V(\phi )
                             \right] \ ,
\end{eqnarray}
where we have chosen the Einstein frame.  
 Since $\beta_i$ at present can be different from
$\beta_i$ in the very early universe, we do not have any constraint on 
$\beta_i$ in the inflationary stage. Of course, ultimately, $\beta_i$
have to approach the observationally allowed values at present.

 In our setup, the preferred frame is selected by the constrained
 vector field $u^\mu$ which violates Lorentz symmetry. 
 In cosmology, there also exists a natural preferred frame,
 the so-called CMB rest frame. 
 As is shown in the Appendix, these two frames are the same practically.
 In this sense, we could have a degeneracy 
 which has not been recognized so far. 
 Once we notice this degeneracy,  it is clear that
 a natural framework to describe the inflationary universe is
  the Lorentz violating scalar-vector-tensor theory of gravity 
  in the sense of (5). 
 
 If $\beta_i =0$, the action (5) is reduced to the conventional one.
 In that case, we have the chaotic inflation for a generic potential $V$.
 For the exponential potential, we have the exact power law inflation.
 Once we switched on $\beta_i$, the Lorentz violating vector affects
 the inflaton dynamics. Hence, our first concern is how the Lorentz 
 violation modifies the picture of the chaotic inflationary scenario.
 Our second aim is to find the exact inflationary scenario with
 the Lorentz violation. Interestingly, we find the exact solutions
 in the absence of the inflaton potential.

\section{Lorentz Violating Chaotic Inflation}

Now, let us consider the chaotic inflationary scenario and clarify 
to what extent the Lorentz violating vector affects the inflationary
scenario. 

In principle, the preferred frame determined by the vector $u^\mu$
can be different from the CMB rest frame. However, alignment of these 
frames had been achieved during the cosmic expansion
 as is explained in the Appendix.
Therefore, let us consider the homogeneous and isotropic spacetime
\begin{eqnarray}
ds^2 = - {\cal N}^2 (t) dt^2 + e^{2\alpha(t)} \delta_{ij} dx^i dx^j \ , 
\end{eqnarray}
where we have included the lapse function ${\cal N}$. The scale of the
universe is determined by $\alpha$. Due to the constraint, 
we have to take 
\begin{eqnarray}
   u^\mu = \left( \frac{1}{\cal N} , 0 ,0 ,0 \right) \ .
\end{eqnarray}
Notice that the spatial isotropy does not allow spatial components
of $u^\mu$. Given these, we can calculate necessary quantities, 
for example, $\nabla_i u^j = \dot{\alpha}/{\cal N} \delta_{i}^j $ 
and other components vanish.  Now, we obtain the action 
\begin{eqnarray}
S = \int dt \frac{1}{\cal N}e^{3\alpha} 
    \left[ -\frac{3}{8\pi G} \left(1+8\pi G \beta \right) \dot{\alpha}^2 
    + \frac{1}{2} \dot{\phi}^2 -{\cal N}^2 V(\phi) \right] 
\end{eqnarray}
where $\beta (\phi) = \beta_1 +3 \beta_2 + \beta_3$.
Note that $\beta_4$ does not contribute to the background dynamics.
One might think the reduced action (8) looks like non-minimally
 coupled scalar field~\cite{Futamase:1987ua,Fakir:1990eg}. 
 However, the structure of equations of motion
 is quite different. In particular, the feature of perturbations is
 completely different. 
 
 Now, let us deduce the equations of motion.
First, we define the dimensionless derivative $Q'$ by
\begin{eqnarray}
  \dot{Q}  =  \frac{dQ}{d\alpha} \frac{d\alpha }{dt} 
            \equiv Q' \frac{d\alpha }{dt}\ .
\end{eqnarray}
Then, the equations of motion are (we set ${\cal N}=1$ after taking 
the variation) 
\begin{eqnarray}
&&  \left( 1 + \frac{1}{8\pi G \beta} \right) H^2 
  = \frac{1}{3} \left[
  \frac{1}{2} \frac{H^2 \phi^{\prime 2}}{\beta} + \frac{V}{\beta} 
                      \right] \\
&&   \left( 1 + \frac{1}{8\pi G \beta} \right)\frac{H'}{H} 
     + \frac{1}{2} \frac{\phi^{\prime 2}}{\beta} + \frac{\beta'}{\beta} =0 \\
&&  \phi'' + \frac{H'}{H} \phi' + 3\phi' + \frac{V_{,\phi}}{H^2}
             + 3 \beta_{,\phi} =0             \ ,        
\end{eqnarray}
where $\beta_{,\phi}$ denotes the derivative with respect to $\phi$.
We have taken  $H=\dot{\alpha}$ as an independent variable.
As is usual with gravity, these three equations are not independent.
Usually, the second one is regarded as a redundant equation.

The above equation changes its property at the critical value $\phi_c$
defined by
\begin{eqnarray}
8\pi G \beta (\phi_c ) =1  \ .
\end{eqnarray}
\begin{figure}[h]
\includegraphics[height=6cm, width=7cm]{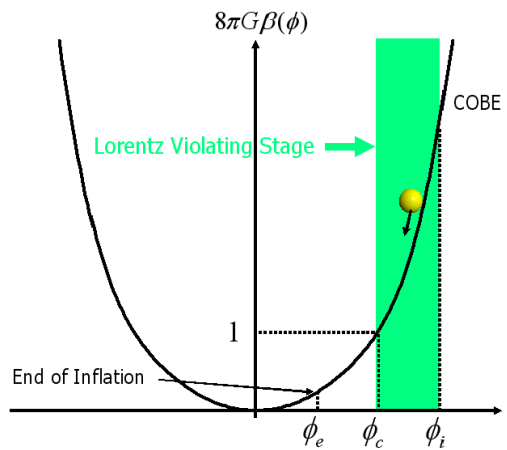}
\caption{There is a critical point where the coupling between
the Lorentz violating vector and the inflaton becomes ineffective.}
\label{fig:1}
\end{figure}
When we consider the inflationary scenario, 
 we usually require the enough e-folding number, say $N=70$. Let $\phi_i$
be the corresponding initial value of the scalar field.
If $\phi_c > \phi_i$, the effect of Lorentz violation on the inflationary
scenario would be negligible. 
However, if $\phi_c < \phi_i$, the standard scenario
should be modified (see Fig.1). It depends on the models. 
To make the discussion more specific, we choose the model
\begin{eqnarray}
  \beta = \xi \phi^2 \ , \quad V = \frac{1}{2} m^2 \phi^2 \ ,
\end{eqnarray}
where $\xi$ and $m$ are parameters. 
For this model, we have 
\begin{eqnarray}
   \phi_c = \frac{M_{pl}}{\sqrt{8\pi \xi}}  \ .
\end{eqnarray}
As $\phi_i \sim 3 M_{pl}$ approximately in the standard case,
 the condition $\phi_i > \phi_c$ implies the criterion
$\xi > 1/(72\pi) \sim 1/226$ for the Lorentz violation to be relevant 
to the inflation. For other models, the similar criterion can
be easily obtained. 

Now, we suppose the Lorentz violation is relevant and
analyze the two regimes separately.

\subsection{Lorentz violating stage}

For a sufficiently larger value of $\phi$, both the coupling function
$\beta$ and the potential function $V$ are important in the model (14). 
During this period, the effect of Lorentz violation on the
inflaton dynamics must be large. 
In the Lorentz violating regime, $8\pi G \beta \gg 1$, we have
\begin{eqnarray}
&&   H^2 
  = \frac{1}{3\beta} \left[
  \frac{1}{2} H^2 \phi^{\prime 2} + V 
                      \right] \\
 &&   \frac{H'}{H} 
     + \frac{1}{2\beta} \phi^{\prime 2} + \frac{\beta'}{\beta} =0 \\         
&&  \phi'' + \frac{H'}{H} \phi' + 3\phi' + \frac{V_{,\phi}}{H^2}
             + 3 \beta_{,\phi} =0     \ .                
\end{eqnarray}
To have the inflation, we impose the condition 
\begin{eqnarray}
H^2 \phi^{\prime 2} \ll V
\end{eqnarray}
as the slow roll condition. Consequently, Eq.(16) is reduced to
\begin{eqnarray}
   H^2   = \frac{1}{3\beta} V   \ .                
\end{eqnarray}
Using Eq.(20), the slow roll condition (19) can be written as
\begin{eqnarray}
\phi^{\prime 2} \ll \beta  \ .
\end{eqnarray}
Now, we also impose the condition $H'/H \ll 1$
 as the quasi-de Sitter condition.
Then, Eq.(17) gives us the condition
\begin{eqnarray}
  \beta' \ll \beta \ .
\end{eqnarray}
We also require the standard condition
\begin{eqnarray}
   \phi'' \ll \phi' \ . 
\end{eqnarray}
Thus, we have the slow roll equations (20) and 
\begin{eqnarray}
  \phi' + \frac{V_{,\phi}}{3 H^2} +  \beta_{,\phi} =0     \ .                
\end{eqnarray}

For our example (14), we can easily solve Eqs.(20) and (24) as
\begin{eqnarray}
   \phi (\alpha) = \phi_i e^{- 4\xi \alpha} \ .
\end{eqnarray}
For this solution to satisfy slow roll conditions (21)$\sim$(23), 
we need $ \xi <1/16$. Thus, we have the range $1/226 <\xi <1/16$
 of the parameter for which the Lorentz violating inflation is relevant.
 Note that, in our model (14), the Hubble parameter (20) becomes constant
\begin{eqnarray}
H^2 = \frac{m^2}{6\xi} \ ,
\end{eqnarray}
even though the inflaton is rolling down the potential.
This is a consequence of Lorentz violation.

\subsection{Standard slow roll stage}

After the inflaton crosses the critical value $\phi_c$, the dynamics
is governed entirely by the potential $V$. 
In the standard slow roll regime $8\pi G \beta \ll 1$, we have
\begin{eqnarray}
&&   H^2 
  = \frac{8\pi G}{3} \left[
  \frac{1}{2} H^2 \phi^{\prime 2} + V  \right] \\
 &&   \frac{H'}{H} + 4\pi G  \phi^{\prime 2}  =0 \\ 
&&  \phi'' + \frac{H'}{H} \phi' + 3\phi' + \frac{V_{,\phi}}{H^2}
            =0                    
\end{eqnarray}
The following arguments are standard. 
The usual slow roll conditions give the slow roll equations
\begin{eqnarray}
&&   H^2 
  = \frac{8\pi G}{3}  V   \\
&&  \phi' + \frac{V_{,\phi}}{3 H^2}  =0     \ .   
\end{eqnarray}
In the simplest case $V= \frac{1}{2} m^2 \phi^2$, 
the evolution of the inflaton can be solved as
\begin{eqnarray}
    \phi^2 (\alpha) = \phi_c^2 - \frac{\alpha}{2\pi G } \ .
\end{eqnarray}
The scale factor $a(t) = e^{\alpha}$ can be also obtained as
\begin{eqnarray}
  a (t)  = \exp \left[ 2\pi G ( \phi_c^2 -\phi^2 (t) ) \right] \ .
\end{eqnarray}
The standard inflation stage ends and the reheating commences
when the slow roll conditions violate.  

\subsection{e-folding number}

Now it is easy to calculate e-folding number.
Let $\phi_i$ be the value of the scalar field corresponding to
the e-folding number $N=70$. 
The total e-folding number reads
\begin{eqnarray}
  N = \frac{1}{4\xi} \log \frac{\phi_i}{\phi_c}
        + 2\pi G \left( \phi_c^2 -\phi_e^2 \right) \ ,
\end{eqnarray}
where $\phi_e \sim 0.3 M_{pl}$ is the value of  
scalar field at the end of inflation.
Note that the first term arises from the Lorentz violating stage.
As an example, let us take the value $\xi = 10^{-2}$. 
Then, $\phi_c \sim 2 M_{pl}$.
The contribution from the inflation end is negligible.
Therefore, we get $\phi_i \sim 12 M_{pl}$. 

In this simple example, the coupling to 
the Lorentz violating sector disappears after 
the reheating. Hence, the subsequent homogeneous dynamics of the universe 
is the same as that of Lorentz invariant theory of gravity.
However, it is possible to add some 
 constants to $\beta_i$, which are consistent with the current
experiments. In that case, the effect of the Lorentz violation
is still relevant to the subsequent history.

So far, we have considered a special model where the coupling $\beta$  
 has the same power as the potential $V$. It is straightforward to
 extend our consideration to more general cases.  

\section{Lorentz Violating Inflation without Potential}

In this section, we will investigate the purely Lorentz violating
inflationary model.

\subsection{Exact Solutions}

It is interesting to observe that we have the inflation even 
in the case $V=0$. In this case, Eqs.(10)$\sim$(12) reads
\begin{eqnarray}
&&   1  = \frac{ \phi^{\prime 2} }{6\bar{\beta}}    \\
&&   \frac{H'}{H} 
     + \frac{1}{2\bar{\beta}} \phi^{\prime 2} 
     + \frac{\bar{\beta}'}{\bar{\beta}} =0 \\
&&  \phi'' + \frac{H'}{H} \phi' + 3\phi' + 3 \bar{\beta}_{,\phi} =0 \ ,   
\end{eqnarray}
where we have defined the variable $\bar{\beta} =\beta + 1/(8\pi G)$.
Substituting Eq.(35) into Eq.(36), we have
\begin{eqnarray}
     ( \bar{\beta} H )' +3 \bar{\beta} H = 0
\end{eqnarray}
It yields $\bar{\beta} H \propto e^{-3\alpha}$.

The condition for the accerelating universe $\ddot{a} >0$ is now
\begin{eqnarray}
    \frac{H'}{H}  > -1 \ .
\end{eqnarray}
Using Eqs.(35) and (36), we can reduce the condition (39)
to 
\begin{eqnarray}
    \left( \log \bar{\beta} \right)' < -2 \ .
\end{eqnarray}
As the scalar is rolling down,
Eq.(35) can be solved as $\phi' = d\phi /d\alpha = -\sqrt{6\bar{\beta}}$.
Thus, finally, we obtain the condition for $\bar{\beta}$ as
\begin{eqnarray}
    \sqrt{\frac{6}{\bar{\beta}}} \frac{d\bar{\beta}}{d\phi} > 2  \ .
\end{eqnarray}

Let us consider an exactly solvable model, $\bar{\beta} = \xi \phi^2$.
In this simplest case, the condition (41) yields $\xi >1/6$.
we can solve Eq.(35) as
\begin{eqnarray}
     \phi \propto e^{-\sqrt{6\xi} \alpha} \ .
\end{eqnarray}
Therefore, we have
\begin{eqnarray}
   H \propto e^{-\alpha /p}  \ , \quad p =\frac{1}{3-2\sqrt{6\xi}}  \ .
\end{eqnarray}
There are three cases to be considered, 
i.e., i)~$1/6 < \xi <3/8$ \ , ii)~$\xi =3/8$ \ , iii)~$3/8 < \xi$ . 

\vskip 0.5cm
\noindent
i)\ \ {$1/6 < \xi <3/8$}\\

In this case, $p>0$. Hence, it is easy to solve Eq.(43) 
\begin{eqnarray}
   \dot{\alpha} \sim e^{-\alpha/p} 
\end{eqnarray}
as
\begin{eqnarray}
    a(t) \sim t^p  \ , \quad p > 1 \ .
\end{eqnarray}
This is a power law inflation. 

\vskip 0.5cm
\noindent
ii)\ \ {$\xi =3/8$}\\

In this case, $1/p =0$. 
This is nothing but the de Sitter solution
\begin{eqnarray}
   a(t) \sim e^{Ht} \ .
\end{eqnarray}
The Hubble constant should be determined by the initial condition. 
Although the spacetime itself is de Sitter,
the scalar field shows the non-trivial time evolution (42).
Therefore, it is interesting to calculate the curvature perturbations
in this model. 

\vskip 0.5cm
\noindent
iii)\ \ {$3/8 < \xi$}\\

In this case, $p\equiv -|p| <0$. Hence, the solution becomes
\begin{eqnarray}
     a(t) \sim (-t)^{-|p|} \ ,   \quad t < 0  \ .
\end{eqnarray}
Thus, this solution represents the super-inflationary universe.
This kind of universe encounters the singularity in the future (at $t=0$).
It is possible to resolve this singularity by adding
the term appeared in the string 
1-loop corrections~\cite{Antoniadis:1993jc,Kawai:1998bn}. 
It is also important to study if the behavior of perturbations
also similar or not~\cite{Kawai:1998ab,Kawai:1999pw}.

\subsection{Inflationary scenario}

In the absence of the inflaton potential,
 we have obtained exact solutions , 
i.e., the power law inflation, the de Sitter inflation, 
and the super-inflation . 
If we slightly modify $\bar{\beta}$, the inflation will end
 when the condition (41) violates.
 Note that, when
the scalar varies from $\phi_i$ to $\phi_e$, 
the e-folding number of the universe can be calculated as
\begin{eqnarray}
   N = \frac{1}{\sqrt{6\xi}} \log\frac{\phi_i}{\phi_e} \ .
\end{eqnarray}
 We expect that the reheating would occur during the oscillation phase.
 It should be stressed that
 the above inflations are completely associated with
 the Lorentz violation.

\section{Evolution of Tensor Perturbations}

Needless to say, it needs to study the evolution of
cosmological perturbations. Due to the Lorentz violation,
the velocity of the gravitational waves are different 
from the velocity of the light.
This and the non-trivial coupling functions $\beta_i$ 
would cause interesting consequence on the spectrum
of tensor, vector  and scalar perturbations.
In particular, the vector perturbations are intriguing
since there are no vector perturbations in the Lorentz
invariant inflationary scenario. However, as the calculation is
very complicated, we leave the complete analysis for future publication.
Instead, here, we discuss the simplest case, namely tensor perturbations.
Even in this case, we can make some interesting predictions.

The tensor part of perturbations can be described by
\begin{eqnarray}
ds^2 =  -dt^2 
    + a^2 (t)\left( \delta_{ij} + h_{ij} (t,x^i) \right) dx^i dx^j  \ ,
\end{eqnarray}
where the perturbation satisfy $h^i{}_i = h_{ij}{}^{,j}=0$.
The quadratic part of the action is given by
\begin{eqnarray}
  S = \int d^4 x \frac{a^3}{16\pi G} \left[
      \frac{1}{4} \gamma
      \dot{h}_{ij} \dot{h}^{\prime ij} 
      -\frac{1}{4a^2} h_{ij,k} h^{ij,k} \right]
\end{eqnarray}
where we have defined 
\begin{eqnarray}
 \gamma = 1- 16 \pi G (\beta_1 + \beta_3 ) \ .
\end{eqnarray}
Apparently, the velocity of the gravitational waves is not 1.
To have the real velocity, we have to impose $\gamma >0$.
Hence, we assume $\beta_1$ and $\beta_3$ are constant.

In the case of chaotic inflation model, the Hubble parameter is constant
 (26) during Lorentz violating stage. The spectrum is completely
flat although the inflaton is rolling down the potential. 
This is a clear prediction of the Lorentz violating chaotic inflation.

This same result applies to the exact de Sitter inflation model
without the potential. In the case of
the power law and the super inflation models, the spectrum of the
primordial gravitational waves are tilted.

\section{Conclusion}

We have examined the impact of Lorentz violation on the inflationary
scenario. As a specific model, we have considered the spontaneous
violation of the Lorentz symmetry due to the vector field.
More specifically, we have investigated scalar-vector-tensor theory
of gravity where the vector is constrained to be unit and time-like.

First, we have examined the chaotic inflationary scenario and found
that the Lorentz violation modifies the dynamics of the inflaton
for a certain parameter region in our model. 
We have shown that the inflationary stage breaks into
 two parts; the Lorentz violating stage and the standard slow roll
 stage. We found that the universe is expanding as an exact de Sitter
 spacetime in the Lorentz violating stage although the inflaton field
 is rolling down the potential. 
Moreover, we have calculated the e-folding number by taking into account
the above modification and shown that we can get enough e-folding number. 

In this paper, we have considered the simplest case 
$\beta \sim V \sim \phi^2$. 
In other cases, for instance, $\beta \sim \phi^4$ and $V\sim \phi^2$,
the Hubble parameter $H$ increases during the Lorentz violating stage.
In the standard slow roll stage, the Hubble parameter $H$ decreases.
Therfore, we can easily generate the spectrum with
the initial (steep) blue spectrum and the later (slightly) red spectrum.
 This may explain the deficiency of the CMB 
 power spectrum at large scales observed by WMAP~\cite{Bennett:2003bz}.

We have also shown that the inflation can be realized
 without the inflaton potential. Depending on the value of
 the parameter $\xi$, we have obtained exact solutions, i.e.
 the power law inflation, de Sitter inflation,
 and the super inflation.  Interestingly,
 even in the exact de Sitter case, the dynamics of the scalar field 
 turns out to be non-trivial.
 In all cases, the inflation ends when the
 coupling function $\bar{\beta}$ is slightly modified from
 exactly solvable case. These exactly solvable models are
 important to understand the evolution of cosmological perturbations
 in the Lorentz violating theory of gravity.
 
To discuss the observability of the effect of Lorentz violation, 
we calculated the tensor perturbations and found the extremely
flat spectrum although the inflaton rolls down the potential
in the case of the chaotic inflation model.
This same result applies to the exact de Sitter inflation model
without the potential. This is a clear prediction of Lorentz violating
inflation model. 

It would be interesting to study the evolution of fluctuation completely.
If the vector modes of perturbations
can survive till the last scattering surface, they leave the remnant
of the Lorentz violation on the CMB polarization spectrum. 
It is also intriguing to seek for a relation to the large scale
anomaly discovered in CMB 
by WMAP~\cite{Eriksen:2003db,deOliveira-Costa:2003pu,Land:2005ad,Copi:2005ff}.
The calculation of the curvature perturbation is much more complicated.
However, it must reveal more interesting phenomena due to
Lorentz violating inflation. The tensor-scalar
ratio of the power spectrum would be also interesting. 
These are now under investigation~\cite{kanno}.

\begin{acknowledgements}
S.K. was supported by JSPS Postdoctoral Fellowships
for Research Abroad. 
J.S. is supported by the Grant-in-Aid for the 21st Century COE "Center for Diversity and Universality in Physics" from the Ministry of Education, Culture, Sports, Science and Technology (MEXT) of Japan, 
the Japan-U.K. Research Cooperative Program, the Japan-France Research
Cooperative Program,  Grant-in-Aid for  Scientific
Research Fund of the Ministry of Education, Science and Culture of Japan 
 No.18540262 and No.17340075.  
\end{acknowledgements}

\appendix

\section{Alignment of preferred frames}

Here, we would like to show the alignment of two frames,
the CMB rest frame and the frame determined by $u^\mu$, 
 will occur during the cosmological
evolution. For simplicity, we ignore the scalar field, 
instead we add the cosmological constant term to the action.
The action is 
\begin{eqnarray}
  S &=& \int d^4 x \sqrt{-g} \left[ \frac{1}{16\pi G} \left(R-2\Lambda \right)
  - \beta_1  \nabla^\mu u^\nu \nabla_\mu u_\nu \right. \nonumber\\
&& \left.  \qquad\qquad\qquad
   -\beta_2 \nabla^\mu u^\nu \nabla_\nu u_\mu    
   -\beta_3  \left( \nabla_\mu u^\mu \right)^2 \right. \nonumber\\
&& \left. \qquad\qquad\qquad
  -\beta_4  u^\mu u^\nu \nabla_\mu u^\alpha \nabla_\nu u_\alpha 
                                   \right. \nonumber \\
&& \left.  \qquad\qquad\qquad + \lambda \left( u^\mu u_\mu +1 \right)
                             \right] \ .
\end{eqnarray}
We consider the Bianchi Type I metric as an ansatz:
\begin{eqnarray}
  ds^2 &=& - {\cal N}^2 (t) dt^2 + e^{2\alpha (t)} \left[ 
  e^{-4\sigma_{+} (t)} dx^2     \right. \nonumber \\
  && \left. 
  + e^{2\sigma_{+} (t) } \left\{ e^{2\sqrt{3}\sigma_{-} (t)} dy^2
  +e^{-2\sqrt{3}\sigma_{-} (t) } dz^2 \right\} \right] \qquad
\end{eqnarray}
and now the vector field can be tilted as
\begin{eqnarray}
  u^\mu = \left( \frac{1}{{\cal N} (t)}\cosh\theta (t), 
  e^{-\alpha (t) +2\sigma_{+}(t)} \sinh \theta (t), 0 , 0  \right) \ .\quad
\end{eqnarray}
Thus, in general, the cosmic frame is different from the preferred frame
determined by $u^\mu$. Substituting the metric and the vector field into
the action (A1), we obtain
\begin{eqnarray}
  S &=& \int dt \frac{1}{\cal N} e^{3\alpha}  \left[ 
    -A \dot{\alpha}^2 -\lambda {\cal N}^2 
    + B \left( \dot{\sigma}_{+}^2 +\dot{\sigma}_{-}^2\right) 
                                                  \right. \nonumber\\
  && \left. \qquad \qquad\qquad
  +D \dot{\theta}^2 -E\theta \dot{\theta} \dot{\alpha}
    -F \theta^2 \dot{\alpha}^2 \right] \ ,
\end{eqnarray}
where
\begin{eqnarray}
A &=& \frac{3}{8\pi G }\left\{ 1
        + 8\pi G \left( \beta_1 +3\beta_2 + \beta_3 \right) \right\} \ , \\
B &=&   \frac{3}{8\pi G }\left\{ 1
         - 16\pi G \left( \beta_1  + \beta_3 \right) \right\} \ , \\
D &=& \beta_1 - \beta_4  \ , \\
E &=& 2\left( 3\beta_2 + \beta_3 +\beta_4 \right) \ , \\
F &=& 2\beta_1 + 9\beta_2 + 3\beta_3 + \beta_4  \ ,\\
\lambda &=& \frac{\Lambda}{8\pi G}  \ .                
\end{eqnarray}
By taking the variation of (A4), we obtain
\begin{eqnarray}
&& A\dot{\alpha}^2 -\lambda = 0  \ , \\
&& \frac{d}{dt} \left( e^{3\alpha} \dot{\sigma}_{\pm} \right) =0 \ , \\
&& A \frac{d}{dt} \left( e^{3\alpha} \dot{\alpha} \right)
                - 3\lambda e^{3\alpha} = 0    \ , \\
&& E \theta \frac{d}{dt} \left( e^{3\alpha} \dot{\alpha} \right)
        - 2D \frac{d}{dt} \left( e^{3\alpha} \dot{\theta} \right) 
        - 2F e^{3\alpha} \theta \dot{\alpha}^2 = 0  \ ,  \quad\qquad     
\end{eqnarray}
where we kept up to the first order with respect to $\sigma_{\pm}$ and $\theta$.
From Eq.(A12), it turns out that the anisotropy decays 
as the universe expands. 
Now, we can deduce the master equation for
the tilt $\theta$ as
\begin{eqnarray}
  \ddot{\theta} + 3\dot{\alpha} \dot{\theta} 
  + \left( \frac{F}{D} \dot{\alpha}^2 
             - \frac{3E}{2AD}\lambda \right) \theta =0   \ .
\end{eqnarray}
Using Eq.(A11) and the definitions (A8) and (A9), we have
\begin{eqnarray}
  \ddot{\theta} + 3\dot{\alpha} \dot{\theta} + \frac{2\lambda}{A} \theta =0 \ .
\end{eqnarray}
For the effective gravitational coupling to be positive, we need $A > 0$.
Thus, (A16) tells us that the tilt $\theta$ 
will vanish during the cosmic expansion.

 Namely, the CMB rest frame and the preferred frame determined by $u^\mu$
are the same practically. What we did in this paper is to
 reveal what this degeneracy means in inflationary cosmology.

\end{document}